\documentclass[11pt, epsf]{article}
\usepackage{epsfig}
\def\ga{\mathrel{\raise.3ex\hbox{$>$\kern-.75em\lower1ex\hbox{$\sim$}}}}
\def\la{\mathrel{\raise.3ex\hbox{$<$\kern-.75em\lower1ex\hbox{$\sim$}}}}

\def\I_M{{I_{\scriptscriptstyle M\times M}}}

\begin{document}

\thispagestyle{empty}
\rightline{IP/BBSR/2003-30}

\vskip 2cm \centerline{ \Large \bf On the Hawking-Page Transition and
the Cardy-Verlinde formula}

\vskip .2cm

\vskip 1.2cm

\centerline{ \bf Anindya Biswas and Sudipta Mukherji}
\vskip 10mm \centerline{ \it Institute of Physics, 
Bhubaneswar-751 005, India} 
\vskip 1.2cm

\centerline{\tt anindyab@iopb.res.in, mukherji@iopb.res.in 
}

\vskip 1.2cm

\begin{quote}

The free energies of the conformal field theories dual to charged adS and
rotating adS black holes show Hawking-Page phase transition. We study the
transition by constructing boundary free energies in terms of order
parameters. This is done by employing Landau's phenomenological theory of
first order phase transition. The Cardy-Verlinde formula is then showed to
follow quite naturally. We further make some general
observations on the Cardy-Verlinde formula and the first order phase
transition.

\end{quote}

\newpage
\setcounter{footnote}{0}


In recent literature, many evidences suggest a
correspondence between gravitational physics in anti-de Sitter (adS) 
space-time and a class of conformal field theories (CFTs) in one lower
dimension \cite{Malda}. This holographic duality, in turn, allowes us to
explore
various space-time physics in terms of field theories which are
non-gravitational in nature. In particular, the thermodynamics of various
black holes, in adS space, is understood as the high-temperature phase of
CFTs. The Hawking-Page (HP) phase transition \cite{Hp} of these black
holes is
then found to correspond to a confining-deconfining phase transition of
${\cal{N}} = 
4$ Yang-Mills gauge theory \cite{Witten} . Furthermore, Verlinde observed
\cite{Verlinde}
that  
for these CFTs, the entropy can be expressed as a Cardy like formula of
two dimensional CFTs \cite{Cardy}. This is now commonly known as
Cardy-Verlinde (CV)
formula. Subsequently, the consequences of having such an entropy  formula
for boundary CFTs were explored in various directions in the
literature \cite{Savver} -  \cite{cno} . In particular, in
\cite{Cm}, the HP transition of the boundary
CFT related to adS-Schwarzschild black hole was studied using Landau
phenomenological theory of first order phase transition. The
Cardy-Verlinde formula was shown to arise from such an analysis in a quite
elegant manner. This motivates us to explore further the relation between
such an entropy formula and the first order phase transition. 

In this letter, we first construct the {\it boundary} free
energies, in terms of the order parameter and the other relevant thermodynamic
variables, for the charged and the rotating adS black holes. For each 
of these black holes, the free energy
is then found to interpolate between two phases around the
HP transition point. Next, we derive the CV
formula from our expressions of free energies. This note ends
with some general observations on the CFTs with first order phase transitions. 
We argue here that the Verlinde's scaling argument, along with the
knowledge of
the temperature and the entropy of the boundary CFT at the phase
transition point, is enough to completely
fix the extensive and the sub-extensive part of the boundary energy.

We start by briefly reviewing the work of \cite{Verlinde} and \cite{Cm} as
this will be useful
for our later purpose. As emphasised in \cite{Verlinde}, for CFT, in a
region with
boundary or in a curved manifold, the energy is not an extensive
quantity. Rather, it has an
extensive part ($E_E$) and a sub-extensive part ($E_c$), often called the
casimir energy. The total energy is then defined as 
\begin{equation}
E = E_E + {1\over 2}E_c, ~{\rm with}~~ E_c = (n-1) (E + pV - TS),
= (n-1)(F +pV),
\label{vdef}
\end{equation}
in $n$ space-time dimensions. Here, $F$ is the free energy. A rather general
scaling argument then suggests that
$E_E$
and $E_c$ depend on the entropy as
\begin{equation}
RE_E = c S^{1 + {1\over (n-1)}}, ~~RE_c = d S^{1 - {1\over (n-1)}}.
\label{eeec}
\end{equation}
Here, $c$ and $d$ are two constants that depend on the detail of the CFT
in question. $R$ is typically the linear size of the system. By
specifying the bulk, one can explicitly determine the constants through
adS/CFT correspondence. Furthermore, by expressing the entropy (using 
(\ref{eeec})) in terms of $E_E$ and $E_c$, one gets the CV formula for
the boundary entropy. We will show this following \cite{Cm}.
We start by considering a $n+1$ dimensional adS
black hole of the form
\begin{equation}
ds^2 = - h(a) dt^2 + {da^2\over{h(a)}} + a^2 d\Omega_{n-1}^2,
\label{bmet}
\end{equation}
with
\begin{equation}
h(a) = 1 - {m\over{a^{n-2}}} + {a^2\over{l^2}}.
\end{equation}
The parameter $m$ is related to the ADM mass $M$ as
\begin{equation}
M = {(n-1) \omega_{n-1} m\over{16\pi g}},
\label{mas}
\end{equation}
where $\omega_{n-1}$ is the volume of the unit $n-1$ dimensional sphere,
 and $g$ is the Newton's constant in $n+1$ dimensions.
The entropy and the temperature are given by
\begin{equation}
{\tilde S} = {a_+^{n-1} \omega_{n-1}\over{4g}}, ~~{\tilde{T}} =
{n\over{4\pi l^2}}
a_+ +
{(n-2) \over{4\pi a_+}},
\label{st}
\label{entt}
\end{equation}
where $a_+$ is the location of the horizon. For large $a$, on a
hypersurface of constant $a$, the metric
behaves as
\begin{equation}
ds^2 = {a^2\over{l^2}} (-dt^2 + l^2 d \Omega_{n-1}^2),
\end{equation}
which is conformally equivalent to  $S^{n-1}\times {\bf R}$ with
$S^{n-1}$ being of radius $l$. More generally, by rescaling
the
time as $\tau = {R\over l} t$, we can rewrite the metric as
\begin{equation}
ds^2 = {a^2\over{R^2}} (-d\tau^2 + R^2 d \Omega_{n-1}^2),
\label{band}
\end{equation}
such that the $S^{n-1}$ is now having radius $R$. This scaling will
enable us to vary the volume of $S^{n-1}$ as we will see later.
The holographic principle then suggests that the theory living on the bulk
which asymptotes to (\ref{band}) is holographically dual to a CFT living
on (\ref{band}) with $R$ being the radius and $\tau$ being the time. 
This prescription, in turn, allows us to identify thermodynamic quantities
of the CFT at high temperature  with the corresponding
thermodynamic quantities of the bulk black hole. 
Following the correspondence, we get the energy ${{E}}$ and
temperature ${{T}}$ of the CFT on $S^{n-1}$ as
\begin{equation}
{{E}} = {M l\over R}, ~~{{T}} = {{\tilde{T}} l\over R},~~S = \tilde S, 
\label{calet}
\end{equation}
where $M$ and $ \tilde T$ are given in (\ref{mas}) and
(\ref{st}) respectively.
From (\ref{vdef}), it is now easy to compute casimir energy $E_c$ and
$E_E$. $E_c$, for example,  is given by
\begin{equation}
E_c = {(n-1)l \omega_{n-1} a_+^{n-2}\over{8\pi g R}}.
\label{cas}
\end{equation}
In deriving $E_c$, we have used the fact that the free energy
is $F = E - T S$ and the pressure is $p = 
-(\partial F/\partial V)_\beta$ with $\beta$ being the 
inverse temperature and $V(= \omega_{n-1} R^{n-1})$ is the volume of $S^{n-1}$.
The equations (\ref{st}), (\ref{calet}) and (\ref{cas}) then
immediately show that the boundary entropy satisfies
\begin{equation}
S = {2\pi R\over {n-1}}{\sqrt{E_c(2 E - E_c)}}.
\label{veren}
\end{equation}
This formula is indeed the Cardy's entropy formula in two dimension once
we identify $RE_c \sim c$ (the central charge) and $RE\sim L_0$ (the zero
mode of the Virasoro generators)\cite{Cardy}. 


The above formula can be re-derived in an elegant manner by exploiting the
first order nature of the HP transition for the adS-Schwarzschild black
hole. Following \cite{Witten}, we can calculate  the {\it boundary} action
for the
bulk metric given in (\ref{bmet}). Without repeating the
calculation, we give the result here:
\begin{equation}
I = {\kappa \omega_{n-1} {\hat a}^{n-1} (1 - \hat a^2)
\over{4 n \hat a^2 + 4(n-2)}}.
\end{equation}
In the above expression, we have written the Newton's constant as $g =
{{l^{n-1}}\over \kappa}$ and introduced dimensionless quantity $\hat a = 
{a_+\over l}$. One then finds the free energy at the {\it boundary} 
$F_{BH} =
\beta^{-1} I$ as
\begin{equation}
F_{BH} = {\kappa \omega_{n-1} \hat a^{n-2} (1 - \hat a^2)\over 
{16 \pi R}},
\label{fre}       
\end{equation}
The boundary field theory undergoes
HP phase transition at a critical value $\hat a =1$. This corresponds to a
critical temperature $T_c$ that can be obtained from  (\ref{st}) and
(\ref{calet}).
Below the critical
temperature, the system prefers pure adS bulk geometry than adS
black hole. While for temperature above $T_c$, it is the black hole phase
which lowers the free energy.

As usual for the first order phase transition \cite{Reichel}, the entropy
changes
discontinuously around the transition point $T = T_c$. 
Following \cite{Cm}, we write the free energy as a function of order
parameter ($\hat a$) and temperature in the following form
\footnote{This expression can be derived following the standard
Landau's phenomenological construction of free energy 
for theories with first order phase transition.
We first make an ansatz,
\begin{eqnarray}
F(\hat a, T) = {\kappa \omega_{n-1}\over R} (p {\hat a}^{n-2}
 - q T {\hat a}^{n-1} + r {\hat a}^n),\nonumber 
\end{eqnarray}
where $p,q,r$ are three constants. $\omega_{n-1}$ is the volume 
of unit $(n-1)$ sphere. The constants can be determined as follows. 
Condition of extremisation, $\partial F/\partial {\hat a} = 0$, must
reproduce the temperature $T$ given in (9). This
determines two of the three constants in terms of the third.
Furthermore, the condition that the 
substitution of $T$ from (9) in $F$ should give (13) determines 
the third constant.}
\begin{equation}
F (\hat a, T) = {1\over 2} E_c(\hat a) ( 1- 2 \beta_c T \hat a + {\hat
a}^2), ~~{\rm where}~~\beta_c = {1\over{T_c}} = {2\pi R\over{n-1}}. 
\label{landau}
\end{equation}
Substituting $E_c$ and $T$ from (\ref{cas}) and
(\ref{calet}) respectively, the above expression for the free-energy
reduces to (\ref{fre}).
We note that for any $E_c$, which is a monotonously growing function of
$\hat a$ with $E_c(0) = 0$, the  free energy (\ref{landau}) describes
a first 
order phase transition from $\hat a =0, ~T < T_c$ to $\hat a > 0, ~T >
T_c$. Furthermore, from the expression of free energy, it follows that the
entropy and the energy are given
by
\begin{eqnarray}
S &=& \beta_c E_c(\hat a) \hat a, \nonumber\\
E &=& {1\over 2} E_c (\hat a) ( 1 + {\hat a}^2).
\label{enten}
\end{eqnarray}
We then have
\begin{equation}
S = \beta_c {\sqrt{E_c (2 E -E_c)}}.
\end{equation}
This is same as  the entropy relation given in (\ref{veren}).

In the next part of the paper, we would like to carry out a similar
exercise
for various other adS black holes. The examples that we will
study here are the charged adS black hole and the rotating adS black hole.
All these black holes show HP transition within certain range of parameters. 
However, due to the presence of various other thermodynamic
potentials, the construction of free energies as a function
of order parameters in the Landau's phenomenological frame work, is 
somewhat involved. We will elaborate on these constrctions in the next 
section of the paper. We will also see that for all these
cases, CV formula appears in a quite natural manner.
\\

\noindent{\bf Charged adS black holes:} The  metric of a charged black
hole in $(n+1)$ adS(RNadS)
space-time is given by
\begin{equation}
ds^2 = -h(a) dt^2 + {da^2\over{h(a)}} + a^2 d\Omega_{n-1}^2,
\label{met}
\end{equation}
with
\begin{equation}
h(a) = 1 - {m\over{a^{n-2}}} + {q^2
\over { a^{2 n -4}}} + {a^2\over{l^2}}.
\label{component}
\end{equation}
The parameters $m$ and $q$ are related to  ADM mass $M$ and charge
$\tilde Q$
as
\begin{eqnarray}
M &=& {(n-1) \omega_{n-1} m\over{16 \pi g}},\nonumber\\
\tilde Q &=& {{\sqrt{2 (n-1)(n-2)}} \omega_{n-1} q\over{8\pi g}}.
\label{mq}
\end{eqnarray}
Here, $\omega_{n-1}$ is again the volume of the unit $n-1$ sphere.
The gauge potential is given by
\begin{equation}
A_t = - {q\over{\alpha a^{n-2}}} + \tilde \Phi, ~{\rm with}~
\alpha = {\sqrt{2 (n-2)\over{n-1}}}.
\end{equation}
In the above equation, $\tilde \Phi$ is a constant. We work with
$\tilde \Phi$
such that the gauge potential at the horizon $a = a_+$ is zero. Hence,
\begin{equation}
\tilde \Phi = {q\over{\alpha a_+^{n-2}}}
\label{const}
\end{equation}
At the horizon of the black hole $h(a_+) = 0$. This condition can be
re-written as
\begin{equation}
m = a_+^{n-2} + {q^2\over{a_+^{n-2}}} + {a_+^n\over{l^2}}.
\label{rmas}
\end{equation}
This expression will be of use later.
The temperature and the entropy associated with the configuration are 
given by
\begin{equation}
{\tilde T} = {n \over{4\pi l^2}}a_+ + {(n-2) (1 - \alpha^2
{\tilde\Phi}^2)\over{4 \pi
a_+}},~~\tilde S = {\omega_{n-1} a_+^{n-1}\over{4 g}}.
\label{temp}
\end{equation}
The on-shell action at fixed gauge potential was calculated for example in
\cite{Chamblin} and is given by
\begin{equation}
I = {\omega_{n-1} \tilde \beta \over{16\pi g l^2}}\Big(l^2a_+^{n-2} 
(1 -\alpha^2
{\tilde\Phi}^2) - a_+^n\Big)
\label{action}
\end{equation}
where $\tilde \beta$ is the inverse of the temperature given in (\ref{temp}).
The grand-canonical potential ${\tilde {G}}({\tilde
T},\tilde\Phi)$ is then given by
\begin{equation}
{{\tilde G}}({\tilde {T}}, \tilde\Phi) = {I\over \tilde \beta} = 
{\omega_{n-1} (1 -\alpha^2{\tilde\Phi}^2) a_+^{n-2}\over{16 \pi g}} - 
{\omega_{n-1}
a_+^n\over{16\pi g
l^2}}.
\label{gibbs}
\end{equation}
Following the adS/CFT prescription,  $n-1$ dimensional CFT on the
{\it boundary} has the temperature, chemical potential, entropy and energy 
\begin{equation}
{{T}} = {l\over R} {\tilde{T}},~~Q = \tilde Q, ~~\Phi = {l\over R} \tilde
\Phi,~~S = \tilde S,~~E = {l\over R}M.
\label{btemp}
\end{equation}
and the Gibbs potential is then
\begin{equation}
{{G}} =  {\omega_{n-1} l(1
-\alpha^2{\tilde\Phi}^2) a_+^{n-2}\over{16 \pi g R}} - {\omega_{n-1} a_+^n
\over{16\pi
g l R}}.
\label{bgibbs}
\end{equation}
Let us now consider ${{G}}$ at a
fixed $\tilde\Phi <1/\alpha$. It is
straightforward to check that ${{G}}$ is negative for 
\begin{equation}
a_+ > l {\sqrt{1 - \alpha^2 {\tilde\Phi}^2}}.
\label{restr}
\end{equation}
Defining $a_c = l {\sqrt{1 - \alpha^2 {\tilde\Phi}^2}}$, we see that the
black
hole is
stable for $a_+ > a_c$ and for $a_+ < a_c$ the adS is preferred with
constant electric potential $\Phi$ all over. From (\ref{btemp}), we see
that this happens at a critical temperature 
\begin{equation}
{{T}}_c = {(n-1){\sqrt{1- \alpha^2 {\tilde\Phi}^2}}\over{2\pi R}}.
\label{crittemp}
\end{equation}
It is now straightforward to write the Gibbs potential for the boundary
theory as an expansion in
terms of order parameter ($\hat a$ as defined earlier) in the
Landau-Ginzburg framework. 
It is given by\footnote{We skip here the systematic constructional
detail. The prescription
is similar to that of the last footnote and also can be found
in [18].} 
\begin{equation}
{{G}}(\hat a, {{T}}) = {\kappa \omega_{n-1} (n-1) (1 - \alpha^2
{\tilde\Phi}^2)\over{16\pi R}} 
{\hat a}^{n-2} - {\kappa \omega_{n-1} {{T}} \over 4} {\hat a}^{n-1}
+ {\kappa (n-1) \omega_{n-1}\over{16 \pi R}} {\hat a}^n.
\label{gb}
\end{equation}
We may mention here that the condition of extremum of the free energy 
with respect to $\hat a$, that is
\begin{equation}
{\partial G\over{\partial {\hat a}}}\Big|_\Phi = 0,
\label{con}
\end{equation}
gives the temperature $T$ as in (\ref{btemp}).
For ${{T}} = {{T}}_c$, ${{G}}$ has  degenerate minima at 
$\hat a = 0$ and $\hat a = \hat a_c$. When ${{T}} < {{T}}_c$,
${{G}}$ has an absolute minimum at $\hat a =0$. For ${{T}} >
{{T}}_c$, the minimum is determined by the larger solution of 
(\ref{con}). 
The energy at
equilibrium can easily be calculated from (\ref{gb}) as follows. We
first find the Helmholtz free energy ${\cal F} = G - Q\Phi$. Then
the energy $E$ is given by:
\begin{equation}
{{E}}(T,V,\Phi) = {\cal F}(T,V,\Phi) - T\Big({\partial {\cal
F}\over{\partial
T}}\Big)_{V,\Phi} =
{(n-1) \kappa \omega_{n-1} {\hat a}^{n-2}\over{16 \pi R}}(1 +
\alpha^2{\tilde\Phi}^2 +
{\hat a}^2).
\label{ener}
\end{equation}
The energy can be seen to be equal to ${ l M\over R}$ where $M$ is given
in (\ref{mq}) and (\ref{rmas}).
Using the generalisation of the formula for $E_c$ given in 
(\ref{vdef}) as 
\begin{equation}
E_c = (n-1)(E + p V - TS - \Phi Q) = (n-1) (F + pV),
\end{equation}
we get
\begin{equation}
E_c = {\kappa (n-1) {\hat a}^{n-2}\omega_{n-1}\over {8 \pi R}}.
\label{ec}
\end{equation}
Here we have used $p = - ({\partial G/\partial V})_{T,Q}$.
We can therefore  rewrite (\ref{gb}) as
\begin{equation}
G(\hat a, T) = {1\over 2}E_c\Big( (1 - \alpha^2 {\tilde\Phi}^2) - {4 \pi R
T\over
{n-1}} \hat a + {\hat a}^2 \Big).
\label{newgb}
\end{equation}
From this expression, we get the CV formula for RNadS black hole as 
\begin{equation}
S = {2 \pi R\over {n-1}} {\sqrt{E_c\Big(2 (E -E_Q) -E_c\Big)}},
\end{equation}
where $E_Q = Q\Phi/2$ is the zero temperature energy of the CFT which
makes contribution to the free energy.
\\

\noindent{\bf Kerr-adS black hole:} The rotating adS black hole metric in
$n+1$
dimensional space-time is
given by
\begin{eqnarray}
ds^2 &=&-{\Delta_a\over {\rho^2}} \Big[ dt - {b\sin^2\theta \over {1
-{b^2\over{l^2}}}}d\phi\Big]^2 + {\rho^2\over{\Delta_a}}da^2 + 
{\rho^2 d\theta^2 \over{\Delta_\theta}} \nonumber\\
&&+ {\Delta_\theta \sin^2\theta \over {\rho^2}}\Big[ b~dt - {a^2 +
b^2\over
{1 -{b^2\over {l^2}}}} d\phi\Big]^2 + a^2 {\rm
cos}^2\theta d\Omega^2_{n-3},
\label{kads}
\end{eqnarray}
where
\begin{eqnarray}
\Delta_a &=& (a^2 + b^2) (1 + {a^2\over{l^2}}) - 2 m a^{4-n}, \nonumber\\
\Delta_\theta &=& 1 - {b^2\over{l^2}}\cos^2\theta, \nonumber\\
\rho^2 &=& a^2 + b^2 \cos^2\theta.
\label{kdef}
\end{eqnarray}
The parameters $m$ and $b$ are related to the black hole energy and
angular momentum as defined later. Note that the metric for large $a$,
 on a hypersurface of fixed $a$, is a
rotating Einstein universe \cite{Hawking}. The temperature ($\tilde T$),
free
energy ($\tilde F$),
entropy ($\tilde S$), energy ($\tilde E$), angular momentum
($\tilde J$) and angular velocity
($\tilde \Omega$) of the black hole, calculated with respect to the
adS background, are given by \cite{Klemm}
\begin{eqnarray}
{\tilde {T}} &=& {(n-2) (1 + {b^2\over {l^2}}) a_+ + n {a_+^3\over{l^2}} +
(n-4) {b^2\over {a_+}}\over{
4 \pi (a_+^2 + b^2)}}, \nonumber\\
{\tilde{F}} &=& {\kappa \omega_{n-1}\over{16 \pi (1 -{b^2\over
l^2})l^{n-1}}} a_+^{n-4}
(a_+^2 + b^2) (1 - {a_+^2\over{l^2}}), \nonumber\\
\tilde S &=& {\kappa \omega_{n-1} \over{4 {(1 - {b^2\over l^2})l^{n-1}}}}
a_+^{n-3} (a_+^2 + b^2),
\nonumber\\
{\tilde{E}} &=& {(n-1) \kappa \omega_{n-1}\over{16 \pi {(1
-{b^2\over l^2})l^{n-1}}}}
a_+^{n-4}(a_+^2+b^2 ) (1 + {a_+^2\over{l^2}}), \nonumber\\
{\tilde{J}} &=& { \kappa \omega_{n-1} b\over{8\pi
{(1 -{b^2\over l^2})^2l^{n-1}}}} a_+^{n-4}(1 + {a_+^2\over l^2})
(a_+^2 + b^2), \nonumber\\
{{\tilde\Omega}} &=& {b (1 + {a_+^2\over{l^2}})\over{(a_+^2 + b^2)}}.
\label{ktherm}
\end{eqnarray}
Here, $a_+$ corresponds to the location of the horizon of the black hole.
This is found by setting $\Delta_a =0$ in (\ref{kdef}).
We first define  the dimensionless
parameters $\hat b = b/l$ and $\hat a = a_+/l$.
Through out this section, we will work with the rotation parameter $\hat b
< 1$. Note that the free energy changes sign and becomes negative for
${\hat a} >1$
signaling a Hawking-Page phase transition. 
 This happens at a critical temperature 
${\tilde T}_c$ given by
\begin{equation}
{\tilde T}_c = {(n-1) + (n-3) {\hat b}^2\over{2 \pi (1 + {\hat b}^2)}}.
\end{equation}
As before, we can understand such a transition at the {\it boundary} 
in the Landau-Ginzburg framework. The {\it boundary} free energy, in the
grand
canonical ensemble, can be written in terms of order parameter
($\hat a$) as\footnote{This expression of free energy can be
systematically constructed following Landau's prescription. But, we leave
here the constructional detail.}
\begin{eqnarray}
{ F}(\hat a, \Omega, { T})  &=& {(n-3) \kappa \omega_{n-1} {\hat b}^2\over
{16 \pi {(1 -{\hat b}^2)R}}} {\hat a}^{n-4} - {\kappa \omega_{n-1}
{\hat b}^2 {T}\over
{4 (1-{\hat b}^2) }}{\hat a}^{n-3} \nonumber\\
&+& {\kappa \omega_{n-1}\over {16 \pi (1 -{\hat b}^2) R}}
\Big( (n-1) + (n-3) {\hat b}^2\Big){\hat a}^{n-2}
- {\kappa\omega_{n-1}{ T}
\over{4 (1-{\hat b}^2) }} {\hat a}^{n-1} \nonumber\\
&+& {(n-1) \kappa \omega_{n-1}
\over{16 \pi (1-{\hat b}^2) R}} {\hat a}^n.
\label{bkerf}
\end{eqnarray}
Note that we could have eliminated $\hat b$ in favour of $\Omega$. However, the
expression of the free energy becomes more involved. That is why we
preferred to retain $\hat b$ in (\ref{bkerf}).
Note also that, in the above,  the boundary thermodynamic quantities are
written without
tildes. More precisely, boundary angular momentum ($J$), temperature
($T$) are related to $\tilde J$ and $\tilde T$ by a $l/R$ scaling
whereas angular velocity $(\Omega)$ remains same as $\tilde \Omega$. 
After somewhat long but straight forward calculation, it follows that the 
location of the extremum of the free energy as a function of order
parameter $\hat a$   
\begin{figure}[ht]
\epsfxsize=10cm
\centerline{\epsfbox{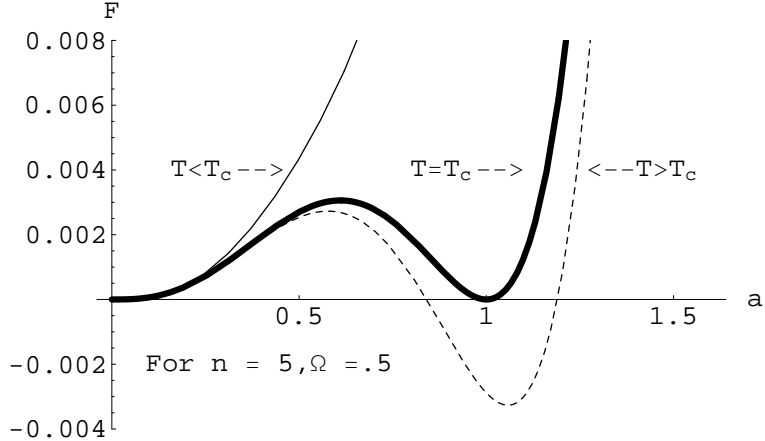}}
\caption{Free energy of the rotating-adS black hole (\ref{bkerf}) as a
function of order parameter $\hat a$ for fixed $R$. Three curves
correspond to three different choices of temperature $T$ as shown.} 
\end{figure}

\begin{equation}
{\partial { F}\over {\partial {\hat a}}}\Big|_\Omega = 0
\end{equation}
corresponds to  the {\it boundary} temperature (${ T}$)
\begin{equation}
{ T} =  {(n-2) (1 + {\hat b}^2) {\hat a} + n {\hat a}^3 +
(n-4) {{\hat b}^2\over
{{\hat a}}}\over{
4 \pi ({\hat a}^2 + {\hat b}^2) R}}.
\label{bt}
\end{equation}
This is nothing but the black hole temperature 
given in (\ref{ktherm}) scaled by appropriate power of $R$. Furthermore,
substituting (\ref{bt}) in to (\ref{bkerf}), we get the
free energy of (\ref{ktherm}), again upto the boundary scaling. It is now
easy to check that ${ F}$ is negative for $\hat a >1$ and is identically
zero at the boundary critical temperature ${ T}_c$, given by
\begin{equation}
{ T}_c = {(n-1) + (n-3) {\hat b}^2\over{2 \pi (1 + {\hat b}^2)R}}.
\end{equation}
Typical boundary phase transition curves are shown in
the figure. We may mention here that the other {\it boundary}
thermodynamic quantities can be
found from (\ref{bkerf}). For example, 
\begin{equation}
J = -{\partial F\over {\partial \Omega}}\Big|_T 
\end{equation}
can easily be evaluated. Consequently, we find that the casimir part of
the boundary energy (\ref{vdef}) can be written as
\begin{eqnarray}
E_c &=& (n-1) (F + pV) \nonumber\\
&=& {(n-1) \kappa \omega_{n-1} {\hat
a}^{n-4} ({\hat a}^2 + {\hat b}^2)\over{8 \pi R (1 -{\hat b}^2)}}.
\end{eqnarray}
Here we have used the thermodynamic relation $p = -(\partial E/\partial
V)_{S,J}$ and $V = \omega_{n-1} R^{n-1}/(1 - {\hat b}^2)$. The CV formula
then follows immediately as 
\begin{equation}
S = {2 \pi R\over {n-1}} {\sqrt{E_c (2 E - E_c)}}.
\end{equation}
\\
{\bf Scaling and phase transition:} We end this note with some
general observations related to finite
temperature CFT and first order phase transition. As we discussed in the
beginning, Verlinde's scaling argument determines the dependence of $RE_c$
and $RE_E$ on the entropy up to some proportionality constants as given in
(\ref{eeec}). These constants, $c$ and $d$, depend on the detail of the
system. We will try to argue here that if the system further shows a first
order phase transition, $c$ and $d$ can be
determined uniquely in terms of critical entropy and temperature.
To begin with, let us consider (\ref{eeec}). These equations can be
inverted to write
\begin{equation} 
S = \Big({RE_c\over d}\Big)^{n\over{n-1}}.
\label{ns}
\end{equation}
Hence the free energy $F = E -TS$ can be written as 
\begin{eqnarray}
F &=& {c\over R} S^{1+ {1\over {n-1}}} + {d\over {2R}} S^{1 -{1\over
{n-1}}} -
TS \nonumber\\
&&\nonumber\\
&=& {1\over 2} E_c\Big( 1 + {2 c\over d}S^{2\over {n-1}} - {2R\over d}T
S^{1\over {n-1}}\Big).
\label{freeexp}
\end{eqnarray}
In writing the above set of equations, we have used (\ref{vdef}).
We will now assume that the system described by (\ref{freeexp})
shows a first order phase transition as we continue to change $T$,
around which the order parameter changes discontinuously (we will continue 
to represent the order parameter by $\hat a$).
We will further {\it assume} that the entropy of the system is a monotonic
function of $\hat a$ with $S =0 ~{\rm at }~ \hat a = 0$. If we now insist
that the free energy vanishes at a non-zero value of $\hat a$ where 
$\partial F/\partial \hat a = 0$, we can then uniquely determine $c$ and
$d$. They are given by 
\begin{equation} 
c = {1\over 2}R T_c S_c^{-{1\over {(n-1)}}}, ~~d = R T_c S_c^{1\over
{(n-1)}}.
\label{cd}
\end{equation}
Here, $S_c$ and $T_c$ are respectively the entropy and the temperature at the
critical point. We therefore have the following scenario. The system shows
a first order phase transition at some critical value of order parameter
at which point $F$ vanishes. However, the system has a finite non-zero
entropy. Below the critical temperature, system prefers a state of zero
order parameter which is chosen to have zero entropy. When the system has
such a phase structure, the extensive and the casimir part of the energy
can
be completely determined in terms of entropy and critical temperature. We
see this by substituting $c$ and $d$ from (\ref{cd}) to (\ref{eeec}):
\begin{equation}
E_E = {T_c\over 2} \Big({S\over S_c}\Big)^{1\over {(n-1)}} S,
~~E_c = T_c \Big({S_c\over S}\Big)^{1\over{( n-1)}} S.
\end{equation}
Note that the explicit dependence of the linear size  of the system
on energies has disappeared in these two expressions.


\noindent{\large\bf Acknowledgments:} 
We would like to thank Somen Bhattacharjee and Goutam Tripathy
for helpful conversations and Biswanath Layek in organizing the figure.
SM would also like to thank Saha Institute
of Nuclear Physics for hospitality where part of the work was done.



\end{document}